\documentstyle{jpsj2}
\def\gs{\gtrsim}
\def\ls{\lesssim}
\def\be{\begin{equation}}
\def\en{\end{equation}}                  
 
\newcommand{\bi}[1]{\mbox{\boldmath$#1$}}
\newcommand{\av}[1]{\langle{#1}\rangle}

\def\bea{\begin{equation}\begin{array}{rcl}}
\def\ena{\end{array}\end{equation}}

\def\q{{\footnotesize{\it q}}\kern -5pt {\footnotesize{\it q}}}
\def\k{{\footnotesize{\it k}}\kern -5pt {\footnotesize{\it k}}}
\def\seq{\sim \kern -12pt \lower 5pt \hbox{$\displaystyle =$}}
\def\nnabla{\nabla\kern-3.3mm\nabla}

\def\ge{> \kern -12pt \lower 5pt \hbox{$\displaystyle =$}}
\def\le{< \kern -12pt \lower 5pt \hbox{$\displaystyle =$}}
\def\gs{> \kern -12pt \lower 5pt \hbox{$\displaystyle{\sim}$}}
\def\ls{< \kern -12pt \lower 5pt \hbox{$\displaystyle{\sim}$}}

\def\be{\begin{equation}}
\def\bea{\begin{eqnarray}}
\def\en{\end{equation}}
\def\ena{\end{eqnarray}}

\newcommand{\tensor}[1]{\stackrel{\leftrightarrow}{\bi{#1}}}
\def\ve{\varepsilon} 
\renewcommand{\theequation}{\arabic{equation}}

\def\runauthor{Akira Onuki}
\begin{document}

\title{Liquid Crystals  in Electric Field }
\author{ Akira  Onuki}
\inst{Department of Physics, Kyoto University, Kyoto 606-8502}
\recdate{\today}

\abst  
{
We present a general theory 
of electric field effects in liquid crystals 
where the dielectric tensor depends 
on the orientation order.  As applications,  
we examine  (i) the director  fluctuations in nematic states 
in electric field for arbitrary strength of the 
dielectric anisotropy  and (ii)  deformation of the nematic order 
around a charged particle. 
Some  predictions are made 
for  these effects. }

\kword{liquid crystals, ions, electric field, 
topological defects}

\sloppy
\maketitle
\section{Introduction}

Phase transitions occur in various systems 
in electric field.  To treat such problems 
we  need to construct a Ginzburg-Landau free 
energy  including  electrostatic interactions 
and electric field effects    under each given 
 boundary condition\cite{NATO}.  
The gross variables include 
the order parameter $\psi$, 
the  polarization $\bi p$, and the  charge 
density $\rho$. (Note $\psi={\bi p}$ in ferroelectric systems.) 
In this letter 
we present such a theory  for liquid crystals in 
nematic states\cite{PGbook}, 
since there seems to be 
no systematic  theory of electric field effects 
for  anisotropic fluids. 
We also  examine 
the deformation of the 
director  around a charged particle, 
since the nematic order around a neutral 
colloid particle or emulsion droplet  has been 
extensively studied\cite{Tere,exp}.  
We shall see that the charge-induced 
orientation is intensified with decreasing 
the particle radius $R$ and/or increasing the charge number $Z$, while 
 the surface anchoring of a  neutral particle 
can be achieved  for large radius because 
 the penalty of the Frank free energy 
needs to be small.

\section{Liquid crystals between capacitor plates}

In liquid crystals the order parameter 
  $\psi$ represents the orientation tensor 
near the isotropic-nematic transition 
or the director $\bi n$ in nematic states. 
We also use the vector notation 
$\bi \rho$ to represent the set 
$\{\rho_\alpha\}$ for the ion densities $\rho_\alpha$ 
($\alpha=1,2, \cdots$).
  We divide the 
total free energy functional  
$F=F_{\rm ch}+F_{\rm st}$ into a chemical part 
$F_{\rm ch}$ and an  electrostatic part $F_{\rm st}$. Here, 
\be 
F_{\rm ch}
= F_0\{\psi, {\bi \rho}\}+ 
\int d{\bi r}  \frac{1}{2}\sum {{\chi}}^{ij} p_ip_j  , 
\label{eq:1}
\en 
The first contribution 
$F_0$ is assumed to be independent 
of  $\bi p$, but there arises  a coupling 
between $\psi$ and $\bi p$ in the presence of 
flexoelectricity\cite{PGbook,Meyer}.  
The tensor $\chi^{ij}$ is the inverse matrix of the 
 the electric susceptibility tensor 
$\chi_{ij}$. We shall see that the local dielectric tensor 
${\ve}_{ij}$ is related to $\chi_{ij}$ as 
\be 
\ve_{ij}=\delta_{ij} +4\pi \chi_{ij}.  
\label{eq:2}
\en 
As is well-known, $\ve_{ij}$ depends on the orientation order. 
In particular, in nematic states the following 
form has been assumed in agreement 
with experiments:\cite{PGbook}  
\be 
\ve_{ij} =\ve_0 \delta_{ij} +\ve_1 n_{i}n_j.   
\label{eq:3}
\en 
The coefficients  $\ve_0$ and $\ve_1$ 
satisfy $\ve_0>1$ and $\ve_0+\ve_1>1$ 
because $\chi_{ij}$ should be a positive-definite matrix 
from the thermodynamic stability.

Next we consider the electrostatic part $F_{\rm st}$. 
A  typical experimental geometry 
is  shown in Fig.1a, where we insert our system  between 
two  parallel metallic plates 
with area $S$ and separation distance $L$. 
We assume $S^{1/2} \gg L$ and neglect the effects of 
edge fields.  
The $z$ axis is taken perpendicularly  to the plates.  
Let the average  
surface charge density of the  upper plate be 
$\sigma_{\rm ex}$ and that of the  
lower plate  be 
$-\sigma_{\rm ex}$. 
The total charge on the upper plate is 
${Q}= S\sigma_{\rm ex}$. 
The electric potential $\phi$   satisfies 
$\phi=0$ at the bottom $z=0$ and $\phi=\Phi$ 
at the top $z=L$, where $\Phi$ is the potential difference between 
the two capacitor plates. 
 The  electric induction $\bi D= {\bi E}+ 4\pi {\bi p}$ 
satisfies 
\be 
\nabla \cdot {\bi D}= -\nabla^2\phi + 4\pi \nabla\cdot{\bi p}= 
 4\pi \rho,  
\label{eq:4}
\en 
where $\rho= \sum_\alpha eZ_{\alpha}\rho_\alpha$ 
is the charge density with  $Z_\alpha e$ 
being the charge of the ion  species $\alpha$. 
The boundary conditions at $z=0$ and $L$ are  
$ E_x=E_y=0$ and $D_z= -4\pi\sigma_{\rm ex}$ \cite{Landau4}. 
 With these relations  
$F_{\rm st}$ is of the form,  
\be 
F_{\rm st}=  \int d{\bi r} {\bi E}^2/{8\pi}. 
\label{eq:5}
\en 
In fact, if  infinitesimal deviations $\delta {\bi p}$, $\delta\rho$, 
and $\delta {Q}$ 
are superimposed on ${\bi p}$, $\rho$,  and ${Q}$, 
the incremental change of  
$F_{\rm st}$ is given by  
\be 
\delta F_{\rm st}= \int d{\bi r}[-{\bi E}\cdot{\delta {\bi p}}+
\phi\delta\rho] 
  + \Phi \delta {Q}, 
\label{eq:6}
\en 
where  use is made of  $\delta ({\bi E}^2)= 
-2{\nabla \phi}\cdot\delta{\bi D}-8\pi{\bi E}\cdot\delta{\bi p}= 
-2\nabla (\phi\delta{\bi D})+8\pi(\phi \delta\rho-{\bi E}\cdot\delta{\bi p})$. 
The right hand side of (6) represents 
the increase of the electrostatic free energy 
due to the small variations of $\bi p$, $\rho$, and $Q$.

If we minimize $F$ with respect to $\bi p$ 
for fixed  $\psi$, $\bi \rho$, and $Q$, 
we require $\delta F/\delta {\bi p}={\bi 0}$. Then 
 (1) and (6) give  
\be 
{\bi p}= \tensor{\chi}\cdot{\bi E}, 
\quad   
{\bi D}= \tensor{\ve}\cdot{\bi E}, 
\label{eq:7}
\en 
which indeed yield (2). Then we may rewrite $F$ as  
$F= F_0+ F_{\rm e}$ with  $F_{\rm e}=\int d{\bi r}\sum_{ij}
\chi^{ij}p_ip_j/2+F_{\rm st}$, so  
\be 
 F_{\rm e}= \frac{1}{8\pi}\int d{\bi r}
  {\bi  E}\cdot 
{{\tensor{\ve}}}\cdot{\bi E} = 
\frac{1}{8\pi}\int d{\bi r}
  {\bi  E}\cdot {\bi D} .
\label{eq:8}
\en
Under these relations 
and using $\delta \chi^{ij}= - \sum_{k\ell}
\chi^{ik}\chi^{j\ell}\delta\chi_{k\ell}$  
we obtain 
\be 
\delta F= \delta F_0 +\int d{\bi r}\bigg [ 
\phi\delta\rho- \frac{1}{8\pi} {\bi  E}\cdot 
{\delta {\tensor{\ve}}}\cdot{\bi E}   \bigg ]+ \Phi\delta{Q}.
\label{eq:9}
\en
From (4) and  (7) the electric potential satisfies  
\be 
-\nabla \cdot\tensor{\ve}\cdot\nabla\phi=4\pi\rho.
\label{eq:10}
\en 
Thus $\phi$ and ${\bi E}=-\nabla\phi$ depend on 
the orientation order via (3) and on $\rho$ under each boundary 
condition. In the literature\cite{PGbook}, however, 
the electric field $\bi E$ in (9) has been replaced by 
the average constant field ${\bi E}_0$,  
where  $F=F_0- \ve_1\int d{\bi r}({\bi E}_0\cdot{\bi n})^2/8\pi$
+const.+O($\ve_1^2$).  This is the  first order approximation 
with respect to $\ve_1$ valid for $|\ve_1| \ll \ve_0$. 
If $\ve_1+ \ve_0$ is considerably smaller 
or larger than $\ve_0$,  
there can be  unexplored regimes  of strong 
polarization anisotropy.

So far  the capacitor  charge ${Q}$ is a control 
parameter and the potential difference $\Phi$ is a fluctuating 
quantity dependent on $\bi p$ and $\bi \rho$. 
We may  also control  $\Phi$ 
 by using  (i) a battery at a fixed potential difference 
 or (ii) connecting  another large 
capacitor in parallel to 
the capacitor containing our system 
as  in Fig.1b.   As will be discussed in the 
appendix, the appropriate free energy functional is given by  
the Legendre transformation \cite{Landau4}, 
\be 
G= F - \Phi {Q} . 
\label{eq:11}
\en   
From (9) the incremental change of $G$ 
is obtained as 
\be 
\delta G= \delta F_0 +\int d{\bi r}\bigg [ 
\phi\delta\rho- \frac{1}{8\pi} {\bi  E}\cdot 
{\delta {\tensor{\ve}}}\cdot{\bi E}   \bigg ]- {Q}\delta\Phi. 
\label{eq:12}
\en

\section{Director fluctuations in 
nematic states }
\setcounter{equation}{12}

As a first application of our theory, 
 we apply an electric field $E_0$ 
to a  nematic liquid crystal  considerably 
below the transition without charges ($\rho=0$),  
where  the orientation order is represented by 
the director vector $\bi n$. Here we do not perform  
the expansion with respect to  $\ve_1$.    For simplicity, we 
assume that $\bi n$ is along the $z$ axis on the average 
under the homeotropic boundary condition and $\ve_1>0$. 
If $\ve_1<0$, the orientation along the $z$ axis becomes 
unstable with increasing $E_0$ above a critical 
value $(\propto L^{-1})$\cite{PGbook}.  
Then  $\bi n$  is written as 
${\bi e}_z+\delta {\bi n}$ with 
$\delta n_z= -[(\delta n_x)^2+(\delta n_y)^2]/2$, where 
${\bi e}_z$ is the unit vector along the $z$ axis 
and $\delta {\bi n}$ is the deviation nearly perpendicular to 
the $z$ axis.  By solving  (10) under  (3), we may expand 
the electrostatic free energy $ F_{\rm e}= F-F_0$ 
in powers of $\delta {\bi n}$. 
The fluctuation contributions  on 
the bilinear order   are  written as 
\be 
\Delta F_{\rm e} = 
\frac{\ve_1}{8\pi}E_0^2 \int_{\bi q} 
\bigg [ |\delta {\bi n}({\bi q})|^2 +
 \frac{\ve_1}{\ve_0+\ve_1\hat{q}_z^2} 
|\hat{{\bi q}}\cdot \delta {\bi n}({\bi q})|^2 
\bigg ] , 
\label{eq:13}
\en  
where 
$\hat{\bi q}= q^{-1}{\bi q}$ 
represents the direction of $\bi q$ and  
${\bi n}({\bi q})$ is the Fourier 
transform of $\delta{\bi n}$.  The  
wave number  $|{\bi q}|$ is assumed to be  much larger than 
the inverse system width   $L^{-1}$.  
The large scale fluctuations are omitted in (13).

Expressing $F_0$ in terms of 
the  Frank constants $K_1$, $K_2$, and $K_3$, 
we obtain  the director correlation functions, 
\be 
\frac{1}{k_{\rm B}T}\av{\delta n_i ({\bi q})\delta n_j ({\bi q})^*} 
=    \frac{\delta_{ij} }{r_2}  
+ \frac{q_iq_j}{q_\perp^2}   
 \bigg ( \frac{1}{r_1}-\frac{1}{r_2}  \bigg ),
\label{eq:14}
\en 
where   $i,j=x,y$,  $q_\perp^2=q_x^2+q_y^2$, and 
\be 
r_1= K_3 q_z^2 + K_1 q_\perp^2 
+\ve_1(\ve_0+\ve_1)(\ve_0+\ve_1\hat{q}_z^2)^{-1}E_0^2/4\pi, 
\label{eq:15}
\en 
\be 
r_2= K_3 q_z^2 + K_2 q_\perp^2 + 
 \ve_1E_0^2/4\pi. 
\label{eq:16}
\en  
If $\ve_1>0$, 
the correlation length $\xi$ is given by 
$
\xi =    (4\pi K/ \ve_1)^{1/2}   E_0^{-1}, 
$ 
where $K$ represents the magnitude of the Frank constants 
The coefficient $r_1$  depends on $\hat{\bi q}$ even 
 in  the limit $q\rightarrow 0$, which is not 
the case in the previous literature.   
The scattered light intensity is proportional to 
the following \cite{PGbook},  
\be 
 \frac{\av{|{\bi f}\cdot \tensor{\ve}({\small{\bi q}})\cdot{\bi i}|^2} 
}{k_{\rm B}T \ve_1^2} = 
 \frac{ |{\bi a}|^2}{r_2}  
+\bigg (\frac{1}{r_1}-\frac{1}{r_2} \bigg )
 \frac{|{\bi q}\cdot{\bi a}|^2}{{q_\perp^{2}}}  ,
\label{eq:17}
\en 
where ${\bi i}$ and $\bi f$ represent 
the initial and final polarizations.  The vector 
${\bi a}$ is defined by 
\be 
a_x= i_zf_x+f_zi_x, \quad a_y=i_zf_y+f_zi_y,\quad a_z=0.
\label{eq:18}
\en  
In addition we examined 
anisotropy of the turbidity (form dichroism) 
in a nematic state under electric field in Ref.$[1]$.

\section{Orientation around a charged  particle}
\setcounter{equation}{18}

We place a charged particle with radius $R$ and charge $Ze$ 
in a nematic state, where $\bi n$ 
is aligned along the $z$ axis or 
${\bi n} \rightarrow {\bi e}_z$  far from 
the particle. Let the density of such charged particles 
be  very low and its Coulomb potential be  not screened 
over a long distance $\lambda$. 
From (3) and (9) the free energy change  
 due to the orientation change ${\delta}{\bi n}$ 
is given by 
\be 
\delta F= -\int d{\bi r} \bigg [ K \nabla^2{\bi n} 
+ \frac{\ve_1}{4\pi}({\bi E}\cdot{\bi n}){\bi E} 
\bigg ]\cdot{\delta}{\bi n} ,
\label{eq:19}
\en 
where we have assumed the 
single Frank constant $K$ ($K_1=K_2=K_3=K$), 
so  $F_0= \int d{\bi r}K|\nabla {\bi n}|^2/2$.  
If the coefficient $\ve_1$ is 
considerably smaller than $\ve_0$, the 
electric field ${\bi E}$ near the particle is  
of the form $-(Ze/\ve_0 r^{2}) \hat{{\bi r}}$ 
and the  electrostatic energy $F_{\rm e}$ is 
approximated by 
\be 
 F_{\rm e}\cong  -
 \ve_1\frac{ Z^2e^2}{8\pi\ve_0^2} 
\int_{r>R} d{\bi r}\frac{1}{r^4} ({\bi n}\cdot{\hat{\bi r}})^2.  
\label{eq:20}
\en 
The origin of the reference frame is taken at  
the center of the charged particle and 
$\hat{{\bi r}}= r^{-1} {\bi r}$. 
Then, for $\ve_1>0$  (or $\ve_1<0$),  
$\bi n$ tends to be parallel 
(or perpendicular) to 
 $\hat{\bi r}$ near the charged particle. 
For $\ve_1>0$ the  decrease of the 
electrostatic energy  is  estimated as   
\be 
 \Delta F_{\rm e}\cong  -\ve_1
\frac{ Z^2e^2}{2\ve_0^2} 
\bigg (\frac{1}{R} -\frac{1}{\ell} \bigg) , 
\label{eq:21}
\en  
where  the radius $R$ plays
 the role of the lower cutoff length and  
 the upper cutoff length $\ell$ is assumed 
to be longer than $R$ and 
shorter than the screening length $\lambda$. 
For $\ve_1<0$, $\ve_1$ in (21) 
should be replaced by $|\ve_1|/3$ (because the 
angle  average of $({\bi e}_z\cdot{\hat{\bi r}})^2$ 
is $1/3$).  In the literature of  physical chemistry,  
much  attention has been paid to 
the decrease of the electrostatic energy 
 around a charged particle  
due to polarization in the surrounding  fluid (solvation 
free energy). 
Its first theoretical expression is the Born formula  
$\Delta E= Z^2e^2(1/\ve_0-1)/2R$\cite{NATO,Born,Is}. 
The additional  orientation degrees of 
 freedom  in  liquid crystals 
  yield    (21).

We are assuming that $\bi n$ is appreciably distorted from 
${\bi e}_z$ in the space region $R\ls r\ls \ell$. 
The  Frank   free energy is estimated as 
\be 
 F_0 \sim  \pi K \ell .
\label{eq:22}
\en 
We determine $\ell$ by minimizing $F_0+ \Delta F_{\rm e}$ to obtain 
\bea 
\ell &=& (|\ve_1|/{2\pi\ve_0^2K})^{1/2}Ze \nonumber\\
&=&(|\ve_1|/{2\pi\ve_0)^{1/2} (a\ell_{\rm B}})^{1/2}Z.
\label{eq:23}
\ena 
Here  $\ell_{\rm B}= e^2/\ve_0T$ is the  Bjerrum length 
and we have set $K= k_{\rm B}T/a$ with $a$ being a microscopic length. 
The condition of strong orientation deformation 
is given by  $R<\ell$ or  
\be 
R/Z<  (|\ve_1|/{2\pi\ve_0)^{1/2} (a\ell_{\rm B}})^{1/2}.
\label{eq:24}
\en 
If this condition does not hold,  
the effect of the electric field on the nematic order 
becomes weak. 
Here $\ell_{\rm B} \sim 500{\rm{\AA}}$ 
for  $\ve_0 \sim 1$ and $T \sim  300$K. 
Therefore,  for a microscopic radius $R$, 
(24)  can well hold unless $|\ve_1|/{\ve_0}$ is very small. 
For salt with small concentration 
consisting of ions with 
charge $Z_\alpha$ and density 
$\rho_\alpha$ ($\alpha=1, 2, ...)$, 
the    screening length 
is given by the Debye-H$\rm{\ddot u}$ckel 
expression  $\lambda^{-2}  = 4\pi
e^2 \sum_\alpha Z_\alpha^2\rho_\alpha/\ve_0k_{\rm B}T$. 
The criterion (24) is meaningful only for 
$\ell <\lambda$. If $R<\lambda< \ell$, 
the director orientation is deformed 
in the space region $R<r<\lambda$ 
and $1/\ell$ in (21) should be replaced by $1/\lambda$.

To illustrate the deformation of 
$\bi n$ around charged objects in equilibrium, 
we have numerically solved     
\be 
{\bi n} \times \bigg [
K \nabla^2{\bi n} 
+ \frac{\ve_1}{4\pi}({\bi E}\cdot{\bi n}){\bi E} \bigg ] ={\bi 0}, 
\label{eq:25}
\en 
and (10) in two dimensions by 
assuming ${\bi n}= (\cos\theta,\sin\theta)$ 
(or $n_z=0$). A charge  
is placed in the hard-core region $(x^2+y^2)^{1/2}<R$. 
In three dimensions this is the case of 
 an infinitely long charged wire with  radius $R$ 
and charge 
density $\sigma$, in which 
all the quantities 
depend only on $x$ and $y$. 
The solution can be characterized by 
the three normalized quantities, 
$\ve_1/\ve_0$, 
$\sigma^*\equiv  \sigma/(\ve_0K)^{1/2}$, and 
$R/\ell_{\rm B}$. Here we set  $\sigma^*=2.4$ and 
$R/\ell_{\rm B}=2$.  We discretize the space into a $200 \times 200$ 
lattice in units of $\ell_{\rm B}$ under the 
periodic boundary condition in the $x$ direction, 
so the system width is $L= 200 \ell_{\rm B}$. 
The electric potential  vanishes  at $y=0$ and $L$. 
In Fig.2, $\bi n$ tends to be 
parallel to the $y$ axis 
far from the origin  and the spacing between 
the adjacent bars is $2\ell_{\rm B}$.   
 The director is in the radial direction for $\ve_1/\ve_0=0.4$ 
in (a)  and is  perpendicular to $\bi r$ 
for $\ve_1/\ve_0=-0.4$ in (b) near the origin. 
A pair of defects are  aligned 
in the $x$ direction in (a) and in the 
$y$ direction in (b). Similar defect formation 
was numerically realized in two dimensions 
for   a  neutral particle  with 
the surface interaction (26) below \cite{Fukuda,Yamamoto}.

\section{Charged  colloidal suspension }
\setcounter{equation}{25}

The nematic orientation around a  
large  particle without charges\cite{Tere,exp} 
is determined by   
competition of the Frank free energy $F_0$
 and a microscopic  anchoring interaction  
at  the particle  surface expressed as  
\be 
F_{\rm a}= - \frac{1}{2}W_{\rm a} \int dS
 ({\bi n}\cdot{\hat{\bi r}})^2.
\label{eq:26}
\en 
Here  $\int dS$ is the surface integration, 
and  ${\hat{\bi r}}$ is the normal unit vector 
at the  surface.  The degree  of  anchoring 
is represented by  the dimensionless parameter,  
\be 
\mu_{\rm a}= W_{\rm a} R/K.
\label{eq:27}
\en

In charged colloidal suspensions,  
the distortion of $\bi n$ 
due to the surface charge can be more important 
than that due to the anchoring  interaction (26).
Note that (24) is well  satisfied even for large $R$ if the 
ratio $Z/R$ can be  increased with increasing $R$.  
Notice that the ionizable points on the surface 
is proportional to the surface area $4\pi R^2$. 
However,  the problem is very complex,  
because the counterions themselves  
can  induce  large deformation 
of the nematic order  (because of their small size) and  
tend to accumulate near the large particles.

For simplicity 
let  the deformation of 
$\bi n$ around the counterions be weak. 
Furthermore, we assume that  the screening length 
$\lambda$    is  shorter than $R$ and the nematic order 
does not vary on the spacial scale of $\lambda$ except 
for the defect-core regions.@ 
Then $F_{\rm e}$ is approximated from (20) as 
$
F_{\rm e}\cong  - W_{\rm e} \int dS  
({\bi n}\cdot{\hat{\bi r}})^2/2$ 
with 
\be 
W_{\rm e}= {\ve_1Z^2e^2\lambda}/{4\pi\ve_0^2}R^4. 
\label{eq:28}
\en 
This is of the same form as $F_{\rm a}$ in (26) 
and the strength of  anchoring  
is represented by  
$
\mu_{\rm eff}= (W_{\rm a}+W_{\rm e})R/K$. 
Here the  Debye-H$\rm{\ddot u}$ckel screening  
becomes extended in the dilute limit. 
In such cases  and in the  limit of large $Z/R$,  
the electric field around a large particle 
decays on the spatial scale of the Gouy-Chapman length\cite{Netz}, 
\be 
\lambda= k_{\rm B}T/eE_{\rm s}= R^2/\ell_{\rm B} Z, 
\label{eq:29}
\en 
where $E_{\rm s}= Ze/4\pi \ve_0R^2$ is the electric field  
at the surface. If $\ell_{\rm B} Z/R \gg 1$, 
we surely have $\lambda \ll R$.  Then we obtain 
$W_{\rm e}= ({\ve_1k_{\rm B}T}/{4\pi\ve_0})Z/R^2$ and  
\be 
\mu_{\rm eff}=   (W_{\rm a}/K)R +  ({\ve_1k_{\rm B}T}/{4\pi\ve_0}K)Z/R. 
\label{eq:30}
\en 
The strong anchoring condition is now  $|\mu_{\rm eff}|>1$.

\section{Concluding remarks} 

In Section 2, we have presented a scheme of  
  Ginzburg-Landau theory for 
the electric field effects in liquid crystals, 
where the dielectric tensor depends on 
the orientation order as in (3). Generalizations 
to more complex situations 
such as ferroelectric cases 
are straightforward.  
In Section 3, the director fluctuations 
have been  examined in applied electric field 
for arbitrary strength of the  dielectric anisotropy. 
It is desirable if (17) could be confirmed 
by  light scattering experiments in systems 
with not small  $\ve_1/\ve_0$. 
In Sections 4 and 5, 
the distortion of the nematic order around 
a charged particle has been studied for the first time. 
The condition of strong orientation deformation  due to the charge effect 
is given by (24) if the charge is 
not screened in the range $r<\ell$ with $\ell$ 
being defined  by (23). 
In colloidal suspensions the presence 
of the counterions makes the problem very 
complex, where we have obtained (30) for  
 the case in which the colloid surface 
charges are screened 
within a thin layer with thickness $\lambda \ll R$.

We  finally mention a similar charge effect  
recently predicted for 
a near-critical polar binary mixture\cite{NATO}.  
In such fluids  
the dielectric constant $\ve(c)$ strongly 
depends on the composition $c$ 
and, as a result, 
 the phase separation behavior 
is  strongly affected by doped ions.  
In particular, if  charged colloid particles 
are suspended, a  charge-density-wave 
phase should be realized for 
$(|\ve_1| /4\ve_0)Z/R> (\pi C_0/\ell_{\rm B})^{1/2}$ 
at low temperatures, 
where $\ve_1= \partial \ve(c)/\partial c$ 
and $C_0$ is a microscopic wave number($=$the coefficient 
in the gradient free energy). Here  large size of 
$(|\ve_1| /\ve_0)Z/R$ is required as in   (24).

This work is supported by Grants in Aid for Scientific 
Research from the Ministry of Education, 
Science, Sports and Culture of Japan.

\vspace{2mm} 
{\bf Appendix}\\
\setcounter{equation}{0}
\renewcommand{\theequation}{A.\arabic{equation}}
We connect two capacitors in parallel as  in Fig.1b.
 One contains an inhomogeneous dielectric material under investigation 
and the other is a large 
capacitor  serving as a charge reservoir.  
The  area $S_0$ and the charge ${Q}_0$ of the large capacitor 
are  much larger  than 
$S$ and ${Q}$ of the smaller capacitor, respectively.  
We are supposing  an experiment in which 
 the total charge ${Q}_{\rm tot}={Q}_0+{Q}$ is fixed and the 
potential difference  
is commonly given by 
$\Phi=  {Q}_0/C_0$, where $C_0$ is the capacitance 
of the large capacitor. Obviously, in 
the limit   ${Q}/{Q}_0 \sim S/S_0 \rightarrow 0$, 
the deviation of $\Phi$ from the   upper bound  
$\Phi_{\rm tot}={Q}_{\rm tot}/C_0$ becomes  
negligible.  Because the  electrostatic energy 
of the large capacitor is given by 
$E_0 = {Q}_0^2/2C_0= ({Q}_{\rm tot}^2/2C_0)  
(1-{Q}/{Q}_{\rm tot})^2$, we obtain   
\be 
E_0 \cong   ({Q}_{\rm tot}^2/2C_0)-\Phi {Q}, 
\label{eq:A.1}
 \en 
where the first term is constant 
and a term of order $({Q}/{Q}_{\rm tot})^2$ is neglected.     
Therefore, for  the total system  including 
the two capacitors, the relevant free energy is given by 
$G$ in (11).


\begin{figure}[b]
\begin{center}
\includegraphics[width=12cm]{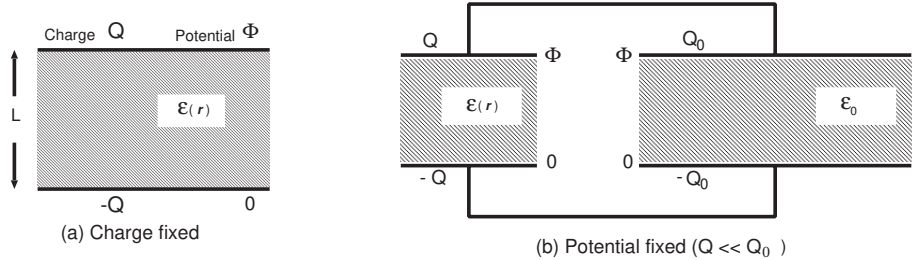}
\end{center}
\caption{(a) System of a capacitor and an 
inhomogeneous  dielectric material  at 
fixed capacitor charge $Q$. The potential difference 
$\Phi$   is a fluctuating 
quantity.  (b) Two 
capacitors connected in parallel with charges ${Q}$ 
and ${Q}_0$. 
 The smaller one  contains  
an inhomogeneous dielectric material,  
and the  larger one 
a homogeneous dielectric material. 
In the limit  ${Q}/{Q}_0 \rightarrow 0$, 
the potential difference $\Phi$ becomes fixed, 
while $Q$ is a fluctuating quantity. }
\label{f1}
\end{figure}

\begin{figure}[b]
\begin{center}
\includegraphics[width=14cm]{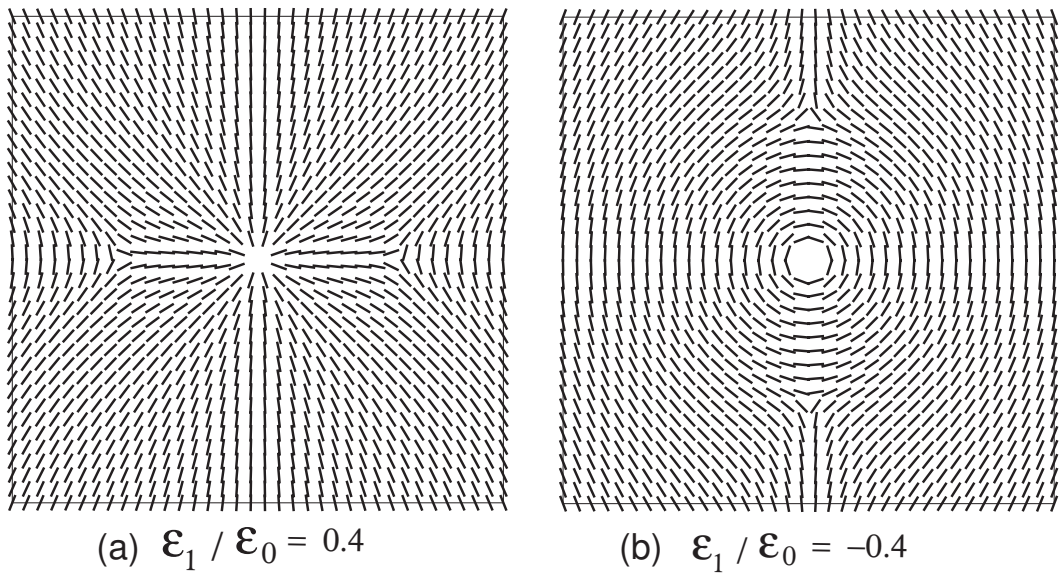}
\end{center}
\caption{
The director in two dimensions 
for $\ve_1/\ve_0=0.4$ 
in (a)  and 
for $\ve_1/\ve_0=-0.4$ in (b) near the origin. 
It  tends to be along  the $y$ axis (in  
the vertical direction) 
far from the origin. The spacing between 
the adjacent bars is $2\ell_{\rm B}$.   
}
\label{f2}
\end{figure}
\end{document}